\documentclass[aps,prl,groupedaddress,twocolumn,showpacs,10pt]{revtex4-1}
\usepackage{graphicx,color}
\usepackage{bm}
\usepackage{amssymb,amsmath,amsfonts, mathrsfs}

\newcommand{\be}{\begin{equation}}
\newcommand{\ee}{\end{equation}}
\newcommand{\ba}{\begin{array}}
\newcommand{\ea}{\end{array}}
\newcommand{\bqa}{\begin{eqnarray}}
\newcommand{\eqa}{\end{eqnarray}}

\begin{document}


\title{Does the bottomonium counterpart of $X(3872)$ exist?}


\author{Zhi-Yong Zhou}
\email[]{zhouzhy@seu.edu.cn}
\affiliation{School of Physics, Southeast University, Nanjing 211189,
P.~R.~China}
\author{Dian-Yong Chen}
\email[]{chendy@seu.edu.cn}
\affiliation{School of Physics, Southeast University, Nanjing 211189,
P.~R.~China}
\author{Zhiguang Xiao}
\email[Corresponding author, ]{xiaozg@ustc.edu.cn}
\affiliation{Interdisciplinary Center for Theoretical Study, University of Science
and Technology of China, Hefei, Anhui 230026, China}


\date{\today}

\begin{abstract}
A narrow line shape peak  at about 10615 MeV, just above the threshold in the
$B\bar B^*$ channel, which can be regarded as
the signal of bottomonium counterpart of $X(3872)$, $X_b$, is
predicted by using the extended Friedrichs scheme.
Though a virtual state is found at about 10593 MeV in this scheme, we
point out that  the peak is contributed mainly
by  the coupling form factor, which comes from the convolution of
the interaction term and meson wave functions including the one from
$\chi_{b1}(4P)$, but not mainly by the virtual-state pole.  In this
picture, the reason why $X_b$ signal is not observed in the
$\Upsilon\pi^+\pi^-$ and $\Upsilon\pi^+\pi^-\pi^0$ channels can also
be understood. The $\chi_{b1}(4P)$  mass and width are found to be
about 10771 MeV  and 6 MeV, respectively and  a dynamically generated
broad resonance is also found with its mass and width at about 10672
MeV and 78 MeV, respectively. The line shapes of these two states are also affected
by the form factor effect. Thus, this study also emphasizes the
importance of the structure of the wave functions of high radial
excitations in the analysis of the line shapes, and provides a caveat
that some signals may be generated from the structures of the form
factors rather than from poles.

\end{abstract}


\maketitle
 Discoveries of the near-threshold exotic states,
$X(3872)$~\cite{Choi:2003ue},
$Z_c$'s~\cite{Ablikim:2013mio,Ablikim:2013emm}, and
$Z_b$'s~\cite{Belle:2011aa}, especially the extremely narrow
$X(3872)$, challenge the predictions of the quark model and different
models  are proposed to understand  their masses, as reviewed
in Refs.~\cite{Guo:2017jvc,Chen:2016qju,Lebed:2016hpi,ESPOSITO20171}.
Hadronic molecular states of $D\bar D^*$ bounded by the long-range
force of one pion exchange~(OPE), were  the first choice of the
explanation of these states, proposed even before these states were
observed~\cite{tornqvistPhysRevLett.67.556}.  This idea is generalized
from the understanding of the deuteron in the triplet $np$
system~\cite{Weinberg:1962hj,Weinberg:1965zz}.  However, this picture
meets problems in explaining the production process in hard $p\bar p$
collision~\cite{Acosta:2003zx,Abazov:2004kp,Aaij:2011sn,Bignamini:2009sk}
and the large  $\gamma \psi'$ decay rate.  Another promising approach
is to consider the $X(3872)$ as dynamically generated  by coupling the
$\chi_{c1}(2P)$ and the opened $D\bar D^*$
continuum~\cite{Braaten:2007dw,
Coito:2012vf,Zhou:2017dwj,Zhou:2017txt}, which may avoid both
problems, and such a picture has attracted more and more interest in
the community.

In the bottomonium sector, the $Z_c$ counterpart $Z_b$'s have already
been found, but the bottomonium counterpart of $X(3872)$, dubbed the
$X_b$~\cite{Hou:2006it}, is still absent in the experimental
explorations. The searches for the $X_b$ in $\Upsilon\pi^+\pi^-$ by
CMS and ATLAS~\cite{Chatrchyan:2013mea,Aad:2014ama}, and in
$\Upsilon\pi^+\pi^-\pi^0$ by Belle~\cite{He:2014sqj} both gave
negative results.  These results call for a reliable theoretical
explanation, and further suggestions of better searching channels are
helpful to save the experimental efforts.  In the literature, the OPE
mechanism was also used in the bottomonium sector and predicts a
binding energy of about 42 MeV for the $B\bar B^*+B^*\bar B$ system of
$J^{PC}=1^{++}$~\cite{Tornqvist:1993ng,Swanson:2006st}, which means
that the bound state is located at about 10562 MeV. The existence of a
bound state around the $B\bar B^*$ threshold was also qualitatively
predicted by considering the isospin exchange mechanism in
Ref.\cite{Karliner:2015ina}.  In Ref.~\cite{Karliner:2014lta}, the possible
mixing between $\chi_{b1}(3P)$ and $X_b$ is also discussed. These
predictions in the literature all assume the $X_b$ to be a bound state
of $B\bar B^*$ by OPE. However, as stated in the previous paragraph,
experiments exhibit that the  $X(3872)$ contains both $\chi_{c1}(2P)$
and $D\bar D^*$ components.  Thus, to have the same production
mechanism as in $X(3872)$, it is more reasonable to couple
$\chi_{b1}(4P)$, the nearest $1^{++}$ bottomonium state above the $B\bar
B^*$ threshold, to the opened continua $B\bar B^*$ and $B^*\bar B^*$.

Motivated by this consideration, we investigate this problem using the
extended Friedrichs
scheme~\cite{Xiao:2016mon,Xiao:2016wbs,Xiao:2016dsx} proposed by us in
recent years.  The basic idea is as follows. The well-accepted
Godfrey-Isgur~(GI) model~\cite{Godfrey:1985xj} can produce the hadron
spectra very well below the open-flavor thresholds but cannot
describe the states above the threshold well because it does not
include the interactions between the hadron states.  The Friedrichs
model~\cite{Friedrichs:1948} is an exactly solvable model that couples
discrete states and the continuum states, which can be used to take
into account the interactions between hadrons. The
interactions could not only shift the discrete-state pole to the
complex energy plane but also dynamically generate other states. Thus,
using the GI's meson spectra as the inputs and using the widely used quark
pair creation~(QPC) model to describe the interactions, the Friedrichs
model provides a way to incorporate the corrections from formerly
neglected interactions in the GI model.  This scheme was successfully
used in describing the first excited charmonium states with only one
free parameter, $\gamma$, the quark pair creation strength. In
particular, it generates $X(3872)$  at the experimental value
automatically and its wave function can be used to understand the
isospin-breaking effects of $X(3872)$ decaying into $J/\psi\pi^+\pi^-$
and $J/\psi\pi^+\pi^-\pi^0$~\cite{Zhou:2017dwj,Zhou:2017txt}, while the
line shape of $D\bar D^*$ in the $B\to D\bar D^*K$ process could be
reproduced well at the same time.  Here, in parallel to the situation
in $X(3872)$, using the same $\gamma$ parameter as in the charmonium
cases, we couple the $\chi_{b1}(4P)$  to $B\bar B^*+B^*\bar B$ and
$B^*\bar B^*$ continuum states, and a few interesting results are
found as follows. We find that a virtual state is dynamically
generated below the  $B\bar B^*$ threshold and the line shape of the
$B\bar B^*$ scattering  has a peak structure just above the threshold,
which seems to indicate that there is an $X_b$ virtual state
generating a peak structure.  However, by careful analysis, we will
show that this peak structure is not contributed mainly by the virtual
state but by the form factor in the amplitude. It is reasonable that
the form factor which comes from the convolution of the meson wave
functions and interaction terms may have nontrivial structures if
higher radial excitations with several nodes in the wave functions are
included.  This picture may also explain why the peak is not seen in
the OZI suppressed $\Upsilon\pi^+\pi^-$ and $\Upsilon\pi^+\pi^-\pi^0$
decay channels. We also show that  the line shape of another
dynamically generated broad resonance and magnitude of the line shape
of a narrow resonance which is originated from the $\chi_{b1}(4P)$ are
also largely affected by the structures of the form factors. These
results show that when high radial excitations are involved, the
effect of the meson structure is important and may even generate peak
signals.

Let us first introduce the theoretical background. By extending the well-known Friedrichs model~\cite{Friedrichs:1948} to  the angular momentum
eigenstates and restricting it to a specific total angular momentum, the
interaction between a discrete state $|0\rangle$ with a bare energy
eigenvalue $m_0$, and some continuum two-particle states
$|E;n,SL\rangle$, where $E$ is the bare energy eigenvalue in the
center of mass system (cms) and $n, S$ and $L$ denote the species, total
spin and total orbital angular momentum, respectively, can be
expressed as a Hamiltonian as~\cite{Xiao:2016mon,Xiao:2016wbs}
\begin{align}\label{fullHamiltonian}
H=&m_0 |0\rangle\langle 0| +\sum_{n,S,L}
\int_{E_{th,n}}^\infty \mathrm dE\, E |E;n,SL\rangle \langle E;n,SL|
 \,\nonumber\\+&\sum_{n,S,L}\int_{E_{th,n}}^\infty \mathrm{d}E  f^n_{SL}(E)|0\rangle\langle
E;n,SL|+H.c.
\end{align}
where ${E_{th,n}}$ is the threshold energy of the $n$th continuum and
$f^n_{SL}(E)$ represents the coupling form factor of the bare discrete
state and the $n$th continuum state.  The general eigenvalue problem of
Eq.~(\ref{fullHamiltonian}) could be exactly
solved. The bound state, virtual state, and resonance state are
determined by the zero points on different Riemann sheets of the
inverse of 
the resolvent, $\eta(z)$,  defined as
\begin{align}
\eta(z)=z-m_0-\sum_{n,S,L}\int_{E_{th,n}}^\infty\frac{|f^n_{SL}(E)|^2}{z-E}\mathrm{d}E\,.\label{eq:eta-pm}
\end{align}
Every continuum integral will contribute a
discontinuity for the $\eta(z)$ function and doubles the number of
Riemann sheets.
For example, in a two-channel case, there are two thresholds, $a_1$
and $a_2$. The physical region between $a_1$ and $a_2$ is attached to
the second sheet, and the physical region above $a_2$ is attached to
the third Riemann sheet.The zero points of $\eta(z)$ on the
unphysical Riemann sheets will be the poles for the $S$ matrix, which
represent the generalized eigenstates with complex eigenvalues for the full Hamiltonian, which have rigorous mathematical definition in the rigged Hilbert space~\cite{Bohm:1989,Civitarese200441}.
The wave functions of such generalized eigenstates could be explicitly written
down~\cite{Xiao:2016wbs,Xiao:2016mon} and the scattering matrix
element of the initial and final continuum states~(the subscripts $i$
and $f$ include their total spin $S$ and angular momentum $L$) could
be expressed as
\bqa\label{scatteringSmatrix}
S_{fi}(E,E')=\delta(E-E')\Big(\delta_{fi}-2\pi i \frac{ f_i(E){f_f}^*(E)}{\eta^+(E)}\Big),
\eqa
In general, only the poles on
the Riemann sheets closest to the physical region will significantly
contribute to the observables such as the cross sections or scattering
amplitudes.

 A simple method to describe the interaction
between one-meson and two-meson continuum states is the QPC
model~\cite{Micu:1968mk,Blundell:1995ev}, in which the transition
operator $T$ of the $A\to BC$ process is
defined as
\bqa
T&=-3\gamma\sum_m\langle 1 m 1 -m|00\rangle\int \mathrm{d}^3\vec{p_3}\mathrm{d}^3\vec{p_4}\delta^3(\vec{p_3}+\vec{p_4})\nonumber\\
&\mathcal{Y}_1^m(\frac{\vec{p_3}-\vec{p_4}}{2})\chi_{1 -m}^{34}\phi_0^{34}\omega_0^{34}b_3^\dagger(\vec{p_3})d_4^\dagger(\vec{p_4}),
\eqa
describing the process of a quark-antiquark pair being generated by
the $b^\dagger_3$ and $d^\dagger_4$ creation operators from the
vacuum.  $\phi_0^{34}=(u\bar u+d\bar d+s\bar s)/\sqrt{3}$ is the
$SU(3)$ flavor wave function for the quark-antiquark pair.
$\chi^{34}_{1-m}$ and $\omega^{34}_0$ are the spin wave function and
the color wave function, respectively.  $\mathcal Y_1^m$ is the solid
harmonic function. $\gamma$ parametrizes the production strength of
the quark-antiquark pair from the vacuum.
By the standard derivation and partial wave decomposition one can obtain $f_{SL}(E)$, the coupling form factor between $|A\rangle $ and $|BC\rangle$ in the Friedrichs model~\cite{Zhou:2017dwj}.

When the wave functions and the masses of the bare states are given,
the form factor $f_{SL}(E)$ can be obtained, and thus, from
Eq.~(\ref{scatteringSmatrix}), the scattering amplitudes of the
particular channels can be obtained and the bound states, virtual
states or resonant states could be solved from $\eta^{n\mathrm{th}\
\mathrm{sheet}}(z)=0$ on the $n$th Riemann sheet. The $G(z)\equiv
f_i(z)f_f^*(z)$ in Eq.~(\ref{scatteringSmatrix}) will be called
the residue function in the following.

We will use the GI model to supply the wave functions
and the mass of the bare state, because it has been proved globally
successful in predicting meson states below the open-flavor
thresholds.  The predicted mass of $\chi_{b1}(4P)$ by GI is about
10790 MeV. Since the $B \bar B^*$ channel threshold is about 10604 MeV
and the $B^* \bar B^*$ channel opens above 10649 MeV,  it is natural
to conjecture that the $\chi_{b1}(4P)$ state will couple to the open
$B^0 \bar B^{0*}+H.c.$, $B^+ \bar B^{-*}+H.c.$,  and $B^{0*} \bar
B^{0*}+B^{+*} \bar B^{-*}$ channels.

\begin{figure}[t]%
\begin{center}%
\includegraphics[height=40mm]{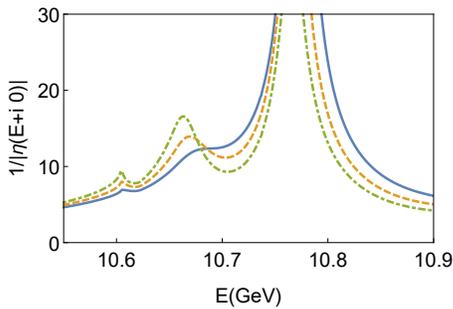}
\caption{\label{fig:recipeta} $|\frac{1}{\eta(E+i 0)}|$ on the real
axis with $\gamma=4.0$(solid line ), 4.9(dashed line), and
5.6(dot-dashed line), respectively.}
\end{center}%
\end{figure}%

The wave functions of $B$, $B^*$, and $\chi_{b1}(4P)$ could be
obtained from the GI model with all the model parameters fixed at
their original values~\cite{Godfrey:1985xj}. Then, the only
undetermined parameter in our calculation, $\gamma$, is chosen at
about $\gamma=4.0$, where the $X(3872)$ and the first excited
charmonium states could be reproduced
simultaneously~\cite{Zhou:2017dwj,Zhou:2017txt}.  By solving the zero
points of $\eta(z)$, three states are found near the physical region:
a virtual-state pole on the second Riemann sheet below the $B\bar B^*$
threshold at $z_v=10593$ MeV,  a pair of conjugate poles at
$z_{R1}=10771\pm 3 i$ MeV on the third Riemann sheet and another pair
of third-sheet conjugate poles at $z_{R2}=10672\pm 39 i$ MeV. From the
curves of $|\frac{1}{\eta({E+i 0})}|$ in Fig.\ref{fig:recipeta}, one
can see that the virtual-state pole contributes a small cusp at the
$B\bar B^*$ threshold, while the other two states contribute two peaks
around the corresponding energies.

\begin{figure}[t]%
\begin{center}%
\includegraphics[height=23mm]{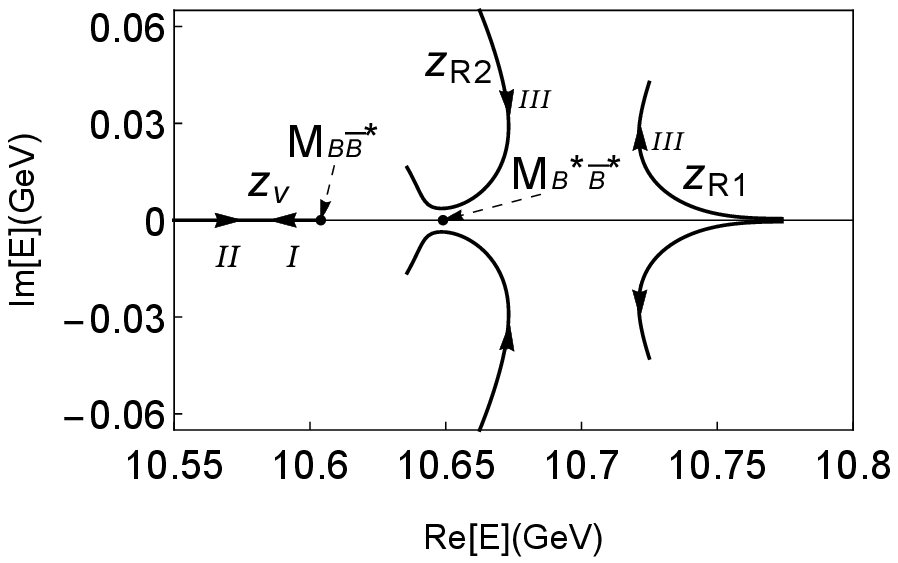}
\hspace{.3cm}
\includegraphics[height=23mm]{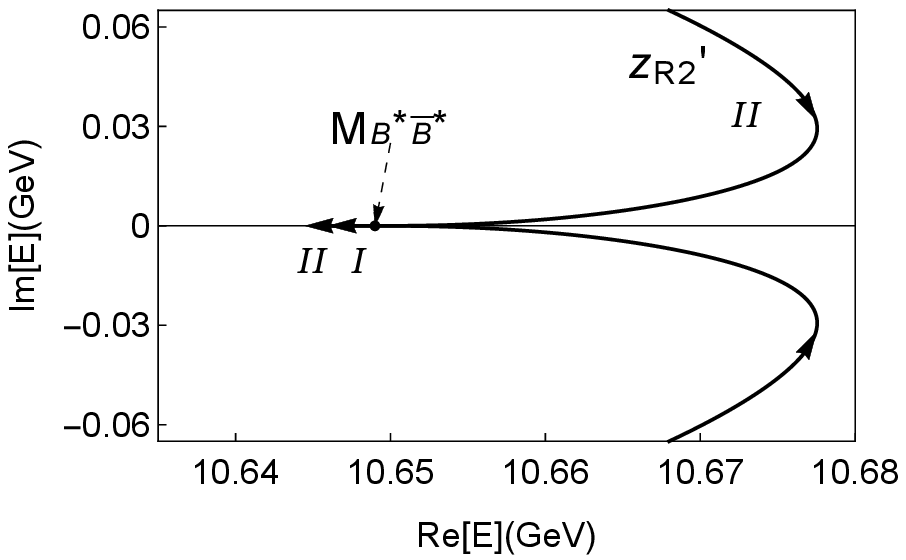}
\caption{\label{fig:poletraj}  The trajectories of  poles on the
complex energy plane as $\gamma$ increases. $I$, $II$, and $III$ denote the sheet numbers of the poles as described in the text. Left: $z_v$, $z_{R1}$, and $z_{R2}$ with both $B\bar B^*$ and $B^*\bar B^*$ continua. Right: $z_{R2}'$ poles with only the $B^*\bar B^*$.  }
\end{center}%
\end{figure}%

The origins of these poles could be
revealed by tracking their pole trajectories along with the change of
$\gamma$, as indicated in Fig.~\ref{fig:poletraj}.  As $\gamma$
becomes smaller, the virtual-state pole $z_v$ moves down towards 
negative infinity on the real axis, remaining as a virtual state.
Conversely, if $\gamma$ is turned larger, the virtual state will move
up along the real axis and reach the threshold at $\gamma\simeq 8.5$,
and then it will come up to the first sheet  becoming a bound state.
This kind of behavior is the typical behavior for dynamically
generated states in $S$ wave with attractive
interaction~\cite{Xiao:2016dsx}.  Therefore, the state can be viewed
as dynamically generated mainly from the  $S$-wave  interaction
between the bare $\chi_{b1}(4P)$ state and $B\bar B^*$ continuum.

 Similarly, one could find that $z_{R1}$ is originated from the
bare $\chi_{c1}(4P)$ state by turning down the $\gamma$ parameter.
$z_{R2}$ is a dynamically generated state mainly from
the coupling between $\chi_{b1}(4P)$ and the $B^*\bar B^*$ continuum.
This can be demonstrated by switching off only the interaction with the
$B\bar B^*$. Then there is only one cut with two Riemann sheets.
As $\gamma$ increases, the $z_{R2}'$
poles, which correspond to the third-sheet $z_{R2}$ poles in the
two-continuum case, will move towards the $B^*\bar B^{*}$ threshold,
 merge at the
threshold, and then become a bound-state pole and a virtual-state
 pole, as shown in the right figure in Fig.~\ref{fig:poletraj}.
 This is a typical
behavior for dynamically generated states in higher partial waves.
Turning on the interaction with $B\bar B^*$ only modifies this
behavior as shown in the left one in Fig.~\ref{fig:poletraj}.
 Therefore, we
conclude that the $z_{R2}$ is mainly generated from the interaction
between $\chi_{b1}(4P)$ and $B^*\bar B^{*}$. The
two dynamically generated states  $z_{R2}$ and $z_v$ behave
differently, because the $S$-wave interaction plays a crucial role in
the formation of $z_v$ and the $D$-wave interaction is responsible for
the formation of $z_{R2}$.

\begin{figure}[t]%
\begin{center}%
\includegraphics[height=25mm]{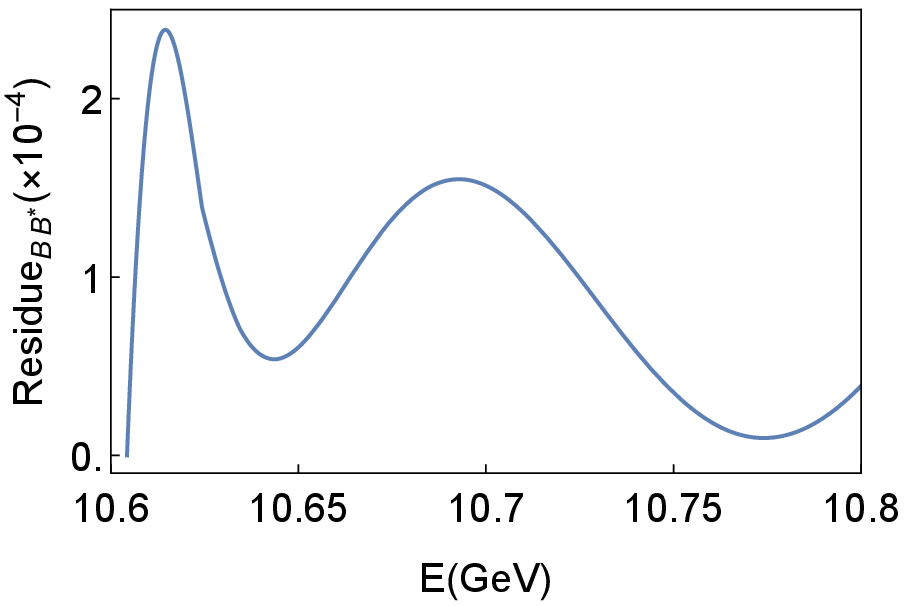}
\hspace{0.7cm}
\includegraphics[height=25mm]{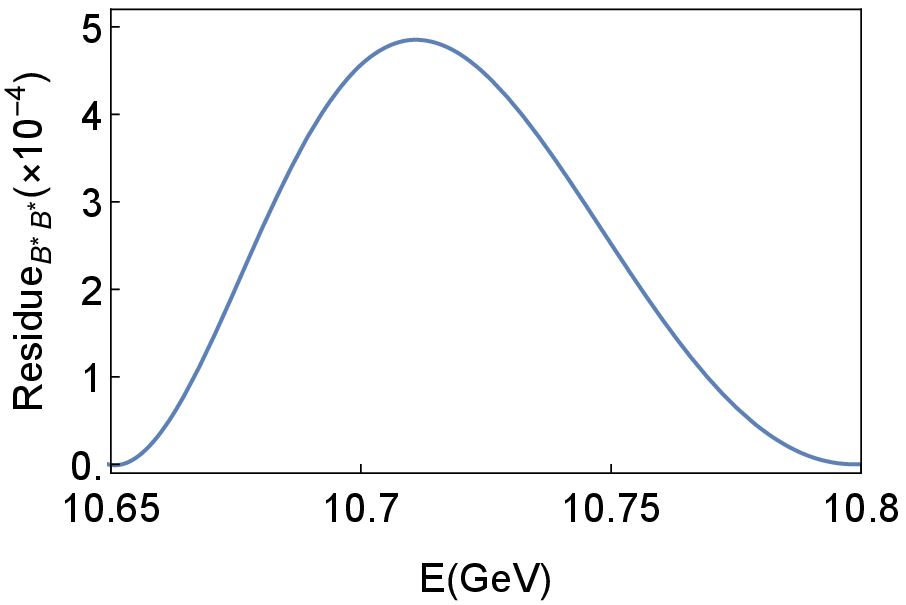}\\
\includegraphics[height=25mm]{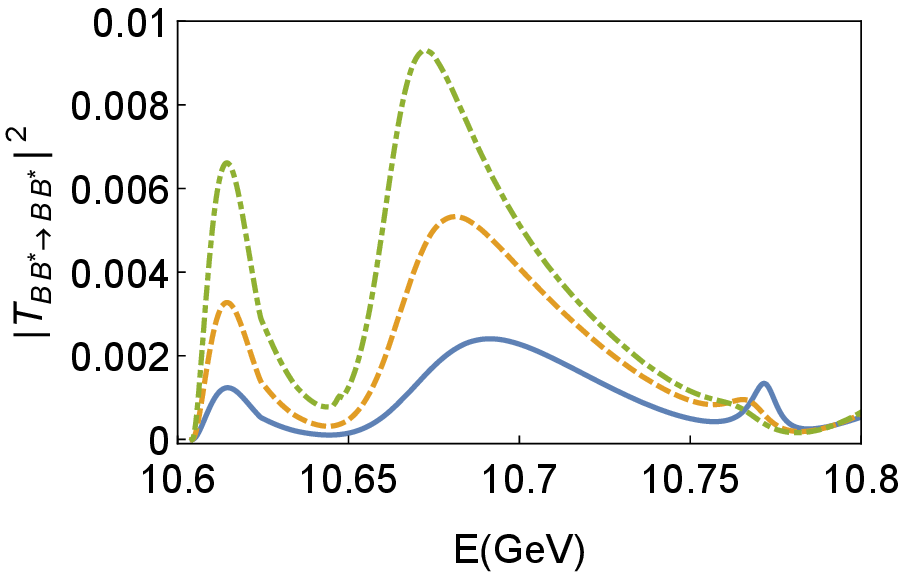}
\hspace{0.1cm}
\includegraphics[height=25mm]{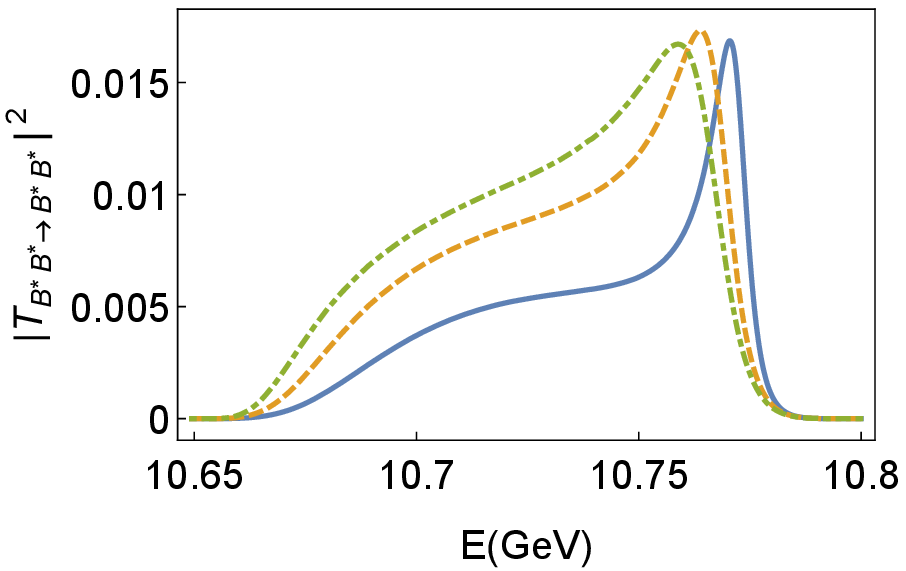}
\caption{\label{fig:TransT}
The residue functions of $B\bar B^*$(left) and $B^*\bar B^*$(right)
are shown in the first row. The absolute square of scattering
amplitudes of $B\bar B^*$ and $B^*\bar B^*$ with $\gamma=4.0$(solid),
4.9(dashed line), and 5.6(dot-dashed line ), respectively, are
shown in the second row.}
\end{center}%
\end{figure}

The physical observables are the cross sections, which are related to
the continuum scattering amplitudes defined in
Eq.~(\ref{scatteringSmatrix}).  The modulus of $B\bar B^*$ scattering
amplitude $|T_{B\bar B^*\rightarrow B\bar B^*}|^2$ exhibits a very
narrow peak in the line shape just above the threshold, as shown in
Fig.~\ref{fig:TransT}.  If this peak is able to be observed in the
experiments and its line shape is crudely fitted with a Breit-Wigner
formula, it will be concluded that there is a state with a mass  about
10615 MeV and a width about 15 MeV by a rough estimation.  However, as
we have shown, there is no such a zero point of the resolvent function
and one can hardly imagine that the virtual state at $z_v$, about $10$
MeV below the threshold on the second sheet, can generate such a
narrow structure near the threshold. In fact, this line shape peak is
mainly contributed by the residue function in
Eq.~(\ref{scatteringSmatrix}), i.e.  the $\sum_{SL}|f^n_{SL}(E)|^2$
terms, but not by the virtual state.  This statement can be clarified
by comparing the  $|1/\eta|$ behavior in Fig.~\ref{fig:recipeta} and
the residue function behavior in Fig.\ref{fig:TransT}. Even when
$\gamma$ is increased to about $5.6$ and the virtual-state pole  moves
to 10600 MeV, its contribution to the threshold enhancement for
$|1/\eta|$ is still not significant as shown in
Fig.~\ref{fig:recipeta}. Only if $\gamma$ is tuned up to 8.0, about
twice of the original 4.0, when the virtual state comes fairly close
to the $B\bar B^*$ threshold, will its contribution be significant.
However, the residue function behavior shown in Fig.\ref{fig:TransT}
presents  a peak just around the one in $|T|^2$.  Since the coupling
function $f^n_{SL}(E)$ in the residue function is obtained from the
QPC model,  it comes from  the convolution of three meson wave
functions and the interactions. For mesons with high radial quantum
number, such as $\chi_{b1}(4P)$ here, it is well known that there
would be several nodes in the radial wave function of the mesons.
Therefore, the wavy structures in the residue functions like in $B\bar
B^*$ scattering in Fig.~\ref{fig:TransT} are closely related to the
structures in the wave functions of high radial excitations. 

In the $B^*\bar B^*$ scattering, $z_{R2}$ and the residue function
together will contribute a bump structure, which also appears in the
$B\bar B^*$ scattering.  The residue function also plays a role in
this structure. The $z_{R2}$ has a broad width of about $78$ MeV at
$\gamma=4.0$, which is expected to be a very  mild structure.   In
$B\bar B^*$ the bump structure around this state is much more obvious
than in $B^*\bar B^*$. This is because its position is in a sharply
rising part of the residue function in $B\bar B^*$, being closer to
the maximum of the bump in the residue function than in $B^*\bar B^*$,
and also because position of the valley of the residue function comes
below $z_{R1}$ in $B\bar B^*$. Thus, this bump structure gets its
shape much more from the residue function than from the pole, which
can easily be seen by comparing Figs.~\ref{fig:recipeta} 
and~\ref{fig:TransT}.  Even though the $z_{R1}$ state is a very
narrow state which receives less influence from the residue function,
a careful observation shows that in the $B\bar B^*$ channel the
position of $\chi_{b1}(4P)$ state is just inside the valley of residue
function, and as a result, its maximum contribution in $|T|^2$ is
comparable to the maximum of the $z_{R2}$ bump, while in  $B^*\bar
B^*$, its peak is rather sharper and higher compared to the mild
structure from $z_{R2}$. This effect is similar to the situation in
the $\Upsilon(5S)$ decay~\cite{TorresRincon:2010fu}

 We also choose another set of meson wave functions, the simple
harmonic oscillator functions with the rms radii from the GI
model, and found that the line shape structures are almost the same
and the above picture is also unchanged.

 In this picture, the absence of the $X_b$ signal in the $\Upsilon\pi\pi$
channel by CMS and ATLAS~\cite{Chatrchyan:2013mea,Aad:2014ama} and in
$\Upsilon\pi\pi\pi$ by Belle~\cite{He:2014sqj} could be understood.
Take the latter for example. If we suppose that the three pions come
from $\omega$~\footnote{A large non-$\omega$ components is found in
Belle's result.} and the lower $\Upsilon\omega$ channel is open but
coupled weakly, the second-sheet virtual pole will move to the third
and fourth sheets, with its mass below the $B\bar B^*$ threshold. This
pole would not affect $1/|\eta|$ significantly in the physical region
which is attached to the second sheet below the $B\bar B^*$ threshold
and to the third sheet above the threshold. The residue function of
the $\chi_{b1}(4P)$ to $\Upsilon\omega$ is OZI suppressed, and thus, any
structure would not be easily observed.  So, we suggest further
experimental searches could pay more attention to the $B\bar B^*$
channel, which could be achieved when the SuperKeKB energy is
increased~\cite{Kou:2018nap}.

In summary, we utilize the extended Friedrichs scheme with the wave
functions and spectrum from the GI model as input to study the pole
structure of $B\bar B^*$ and $B^*\bar B^*$ scatterings by coupling
$\chi_{b1}(4P)$ to $B\bar B^*$ and $B^*\bar B^*$.  It is for the first
time that a line shape peak related to the $X_b$, at about 10615 MeV,
is generated just above the $B\bar B^*$ threshold and at the same time
the $X(3872)$ can be described well in one consistent scheme.  In
comparison, the tetraquark model could predict an $X_b$ state above
the $B\bar B^*$ threshold but  cannot describe the  mass for $X(3872)$
well~\cite{ALI2017123}.  In addition, we have shown that it is the
residue functions in the amplitude, which are related to the wave
functions of the mesons, rather than the dynamically generated virtual
state that contribute to the $X_b$ peak dominantly. Our picture could
also explain the absence of $X_b$ signal evidence in the
experiments~\cite{Chatrchyan:2013mea,He:2014sqj}.  A dynamically
generated resonance pole is also found at $10672\pm 39i$ MeV and the
$\chi_{c1}(4P)$ at $10771\pm 3i $ MeV. The line shapes of these
resonances are also affected by the residue functions. In particular,
the relative magnitude of the peak generated by the narrow resonance
$\chi_{b1}(4P)$ is suppressed by a valley of the residue function.
This kind of phenomenon may be a general cause for a state to behave
differently in different channels when the form factors are different.
In principle, the wave function of a high radial excitation state has a few
node structures, and, after being convoluted with the interaction Hamiltonian
it may finally cause additional structures of the line shape.  This
effect is independent of the model chosen in this paper.  Thus, the 
wave functions with higher radial quantum numbers may result in  more
structures in the line shape. This kind of phenomenon is more or less a 
model-independent one and should be paid attention to in the theoretical
analysis of the line shape data. In comparison, in the effective field
theory (EFT) approach, the states are assumed to have no internal
structures and the information of the form factors is absorbed into
the coupling constants.  Therefore, the EFT approach must go to higher
orders to reproduce the nontrivial behaviors of the form factors or
must include some form factors inserted by hand without any solid
theoretical ground, which may constrain the effectiveness of the
theory.

\begin{acknowledgments}
This work is supported by the Natural Science Foundation of Jiangsu Province
of China under Contract No. BK20171349 and China National Natural Science Foundation under
Contract No.  11105138, No. 11575177, No. 11235010 and No. 11775050.
\end{acknowledgments}

\bibliographystyle{apsrev4-1}
\bibliography{Ref}

\end{document}